\documentclass[11pt]{article}
  
\usepackage{amsmath,amssymb}

\usepackage{epsfig}  
\usepackage{graphicx}               
\usepackage[hyphens]{url}
\usepackage{hyperref}
\usepackage{breakurl}

\usepackage{cite}
\usepackage{bm}
\usepackage{cancel}
\usepackage{pstricks}
\usepackage{color}
\usepackage{braket}
\usepackage{esvect}

\newcommand{\nc}{\newcommand}  

\nc{\beq}{\begin{equation}}  
\nc{\eeq}{\end{equation}}  
\nc{\beqa}{\begin{eqnarray}}  
\nc{\eeqa}{\end{eqnarray}}  
\nc{\bit}{\begin{itemize}}  
\nc{\eit}{\end{itemize}}

\title{
\vspace*{-2.3cm}
\begin{flushright}
\normalsize{
  }
\end{flushright}
\vspace{1.5cm}
\Large  
\textbf{Detecting Axion Stars with Radio Telescopes
} \vspace*{1.0cm}   
}

\author{{\bf Yang Bai}$^a$ and {\bf Yuta Hamada}$^{a,b}$
\vspace{5mm}
\\
$^a$\normalsize\emph{Department of Physics, University of Wisconsin-Madison, Madison, WI 53706, USA}  \vspace{1mm} \\
$^b$\normalsize\emph{KEK Theory Center, IPNS, KEK, Tsukuba, Ibaraki 305-0801, Japan}
}
\date{}

\begin{document}  
\setcounter{page}{0}  
\maketitle  

\vspace*{1cm}  
\begin{abstract} 
When axion stars fly through an astrophysical magnetic background, the axion-to-photon conversion may generate a large electromagnetic radiation power. After including the interference effects of the spacially-extended axion-star source and the macroscopic medium effects, we estimate the radiation power when an axion star meets a neutron star. For a dense axion star with $10^{-13}\,M_\odot$, the radiated power is at the order of $10^{11}\,\mbox{W}\times(100\,\mu\mbox{eV}/m_a)^4\,(B/10^{10}\,\mbox{Gauss})^2$ with $m_a$ as the axion particle mass and $B$ the strength of the neutron star magnetic field. For axion stars occupy a large fraction of dark matter energy density, this encounter event with a transient $\mathcal{O}(0.1\,\mbox{s})$ radio signal may happen  in our galaxy with the averaged source distance of one kiloparsec. The predicted spectral flux density is at the order of $\mu$Jy for a neutron star with $B\sim 10^{13}$~Gauss. The existing Arecibo, GBT, JVLA and FAST and the ongoing SKA radio telescopes have excellent discovery potential of dense axion stars.  
\end{abstract} 
  
\thispagestyle{empty}  
\newpage  
  
\setcounter{page}{1}

\vspace{-2cm}

\section{Introduction}
\label{sec:intro}
The QCD axion, as a byproduct of a solution to the strong CP problem~\cite{Peccei:1977hh,Weinberg:1977ma,Wilczek:1977pj,Shifman:1979if,Kim:1979if,Zhitnitsky:1980tq,Dine:1981rt}, can also serve as a cold dark matter of the universe~\cite{Turner:1989vc}. The axion is the pseudo Nambu-Goldstone boson corresponding to the spontaneous symmetry breaking of the Peccei-Quinn (PQ) symmetry. The PQ symmetry is slightly broken by the chiral anomaly, and hence the axion acquires a tiny mass through the physics at the QCD scale. The mass of the axion can be estimated as $m_af_a\approx f_\pi m_\pi\approx (10^8\,{\rm eV})^2$~\cite{Kaplan:1985dv}, where $f_{a(\pi)}$ and $m_{a(\pi)}$ are the decay constant and mass of the axion(pion). The minimum of the axion potential corresponds to the CP conserving coupling constant $\bar{\theta}=0$, and the strong CP problem is dynamically solved by introducing the new particle. Moreover, in the evolution of the universe, the universe may experience the phase transition of PQ symmetry breaking. At a very high temperature, the axion is essentially massless with all possible values of $\bar{\theta}$ and the axion field takes value which is different from $\bar{\theta}=0$. When the universe cools down to a temperature around the QCD scale, the axion mass sits in and becomes comparable to the Hubble expansion rate. The axion field starts to oscillate around its minimum and behaves as a condensate of axions~\cite{Preskill:1982cy,Abbott:1982af,Dine:1982ah}. This vacuum misalignment mechanism results in the coherent oscillation of axions, which can explain the observed dark matter energy density.\footnote{In addition to the vacuum misalignment, the axion can be produced by the radiation from cosmic string~\cite{Davis:1986xc} and the decay of string-wall system~\cite{Lyth:1991bb}. The numerical simulation~\cite{Kawasaki:2014sqa} favors $m_a\simeq 10^{-4}$eV for $N_{\rm DW}=1$.}

If the PQ symmetry breaking occurs after the end of the inflation, the axion field value takes random value for each Hubble volume. This large spacial fluctuation of the axion field is further enhanced by the gravitational instability. As a result, the axion mini-cluster~\cite{Hogan:1988mp,Kolb:1993zz,Kolb:1995bu} is very likely to be formulated, and it becomes axion star after the gravitational thermalization~\cite{Sikivie:2009qn,Guth:2014hsa}. Depending on the energy density, there are two types of axion star: diluted axion star and dense axion star~\cite{Braaten:2015eeu}. The diluted axion star is formed by the balance between the kinetic pressure and gravitational attractive force, and the dense axion star is formed by the balance between the self interaction of the axion and the gravitational force. For a fixed axion-star mass, the radius of a diluted axion star can be several orders of magnitude larger than the dense one. 

The searches for background axions, not inside an axion star, rely on the axion ``invisible" couplings to the Standard Model (SM) particles, which are suppressed by the PQ scale, $f_a$, above $\sim 10^8$\,GeV from experimental constraints~\cite{Graham:2015ouw}. Because of the suppressed coupling strength, it is challenging, but possible to have table-top experiments to detect axion dark matter particles~\cite{Sikivie:1983ip,Asztalos:2009yp,Barth:2013sma,Armengaud:2014gea,Brubaker:2016ktl}. To search for axion mini-clusters or stars, the gravitational lensing (micro-lensing) experiments could be sensitive for the mass range of QCD axion stars. Recently, the mirco-lensing-based searches for massive astrophysical compact halo object (MACHO) by Subaru Hyper Suprime-Cam in Ref.~\cite{Niikura:2017zjd} has imposed a constraint on the fraction of axion-star energy in total dark matter energy $f_{\rm AS}$ to be below a few percent for $M_{a\odot}=10^{-12}M_\odot$ to no constraint for $M_{a\odot}\approx10^{-14}M_\odot$. Future micro-lensing experiments will further constrain on the fraction of axion stars as a part of dark matter. Another novel way to look for axion stars is to look for the hydrogen axion star, which may well be formed by ordinary baryonic matter becoming gravitationally bound to an axion star~\cite{Bai:2016wpg}.  

In this paper, we focus on the brightness of axions when they pass through some astrophysical environment with a large magnetic field. In the presence of the axion field, the electromagnetism becomes axion electromagnetism~\cite{Wilczek:1987mv}. The axion coupling to two photons via $a\,F_{\mu\nu}\widetilde{F}^{\mu\nu}$ can convert axion to real photons in the magnetic background. If the conversion rate is sufficiently large, we could detect the signal as a monochromatic radio wave signal with a frequency of $\nu=m_a/2\pi$. In this respect, the axion star is the interesting object because it may enhance the signal due to its large energy density or field value at the core. To know the populations of axion stars in our galaxy or Universe, it is important to know the fraction of all axion particles inside axion stars, which requires a more realistic numerical calculation than Ref.~\cite{Khlebnikov:1999qy}. In this paper, we treat the fraction of the axion star as a phenomenological parameter, and investigate the observational consequence. For the axion star mass, we  concentrate on the range of $10^{-14}$--$10^{-12}\,M_{\odot}$, as estimated for the QCD axion~\cite{Guth:2014hsa,Braaten:2015eeu,Bai:2016wpg}. 

We first estimate the radiated power from the homogeneous axion component interacting with the magnetic field in our galaxy/neutron star, which turns out to be too small to be detected by on-going and future radio telescope~\cite{Pshirkov:2007st,Kelley:2017vaa,Sigl:2017sew}. Afterwards, we estimate the radiation power of the axion star in the static magnetic field, $10^{10}$--$10^{15}$\,Gauss, associated with the neutron star, which can be potentially large enough to be detected in the radio telescopes. In our calculation, we properly take account of the two important effects: the {\it interference} effect and {\it medium} effect. At the leading order in perturbation (the source size over its distance to the observer), the radiated field can be written in terms of the integration of the source term multiplied by the retarded Green's function. The integration over the source region leads to the interference effect, and the resultant radiation field is suppressed by powers of $1/(m_a\,R_{\rm AS})$ depending on the geometry of integration range, where $R_{\rm AS}$ is the radius of the axion star. Even for dense axion stars with $R_{\rm AS}\sim 1$\,m, the suppression factor is multiple power of $10^{-3}$. For the second medium effect, we consider the response of the electron-ion plasma, the state in the outer crust of neutron stars, to the axion-generated electric field. Using a simple Drude model, we calculate the electric susceptibility of the plasma medium and then the radiation power based on the macroscopical Maxwell equations. We have found that the power is not strong enough to explain the fast radio burst (FRB)~\cite{Lorimer:2007qn}, contrary to the claim in Refs.~\cite{Iwazaki:2014wta,Iwazaki:2015zpb,Raby:2016deh}, where the above effects are not included. On the other hand and for a neutron star with a large magnetic field, the radiation power could be large enough to be detected by the on-going or future radio telescopes such as  Arecibo~\cite{Giovanelli:2005ee}, GBT~\cite{GBT}, JVLA~\cite{JVLA}, FAST~\cite{Nan:2011um} and SKA~\cite{SKA}.

This paper is organized as follows. In Section~\ref{sec:axion-stars}, we review the dilute and dense axion star properties. In Section~\ref{sec:axion-EM}, we estimate the radiation power from homogeneous axion dark matter in our galaxy. In Section~\ref{sec:radiation}, we calculate the radiation power of the axion star in the static magnetic field associated with the neutron star. In Section~\ref{sec:neutron-star}, we derive the encounter rate of the axion star with the neutron star, and investigate the detectability of the current and future observations. We then conclude in Section~\ref{sec:conclusion}. 

\section{Axion-star Mass and Size}
\label{sec:axion-stars}
Depending on the axion star mass, $M_{a\odot}$, and other axion model parameters: $m_a$ and $f_a$, one may have diluted or dense axion stars as components of dark matter, if the PQ-symmetry breaking scale is below the end of inflation scale $H^{\rm end}_{\rm Inflation} \lesssim 10^{13}$~GeV~\cite{Ade:2015lrj}.  For the diluted axion star, the axion self-interactions usually can be ignored. As studied in the literature, this requires the axion star mass to satisfy~\cite{Chavanis:2011zm}~\footnote{See Refs.~\cite{Eby:2016cnq,Braaten:2016dlp,Eby:2017xrr} for recent discussion about the axion-star stability.}
\beqa
M_{a\odot} = {\cal N}_a \,m_a \lesssim 10.15~f_a/(m_a \, G_N^{1/2}) = 1.0 \times 10^{-13}\,M_{\odot}\times\,\left( \frac{f_a}{10^{11}~\mbox{GeV}} \right)\,\left( \frac{10^{-4}~\mbox{eV}}{m_a} \right)
\,,
\label{eq:dilute-condition}
\eeqa
with $G_N$ as the Newton constant. For the QCD axion, we anticipate $m_a f_a \approx m_\pi f_\pi \approx (10^8\,\mbox{eV})^2$~\cite{Kaplan:1985dv}. To account for all dark matter energy density, the axion mass is $10^{-5}$--$10^{-3}$~eV~\cite{Turner:1989vc}. We therefore choose $m_a=10^{-4}$~eV and $f_a=10^{11}$~GeV as a benchmark model. Using a variation calculation based on the hydrogen $1S$ wave function, it is easy to show that the radius of the diluted axion star is 
\beqa
R^{\rm dilut}_{\rm AS} \sim \frac{1}{ M_{a\odot}\, G_N \, m_a^2} \approx 2.7\times 10^{6}\,\mbox{m}\times 
\left( \frac{10^{-15}\,M_{\odot}}{M_{a\odot}} \right)\,\left( \frac{10^{-4}~\mbox{eV}}{m_a} \right)^2
\,.
\label{eq:dilute-axion-size}
\eeqa
Its energy density in the unit of $m_a^2 f_a^2$ is very small and has $\rho_{a\odot}/m_a^2 \,f_a^2 \approx  4.5\times 10^{-21}$ for the same model parameters of the above equation. 
%
%

\begin{figure}[th!]
\begin{center}
\includegraphics[width=0.6\textwidth]{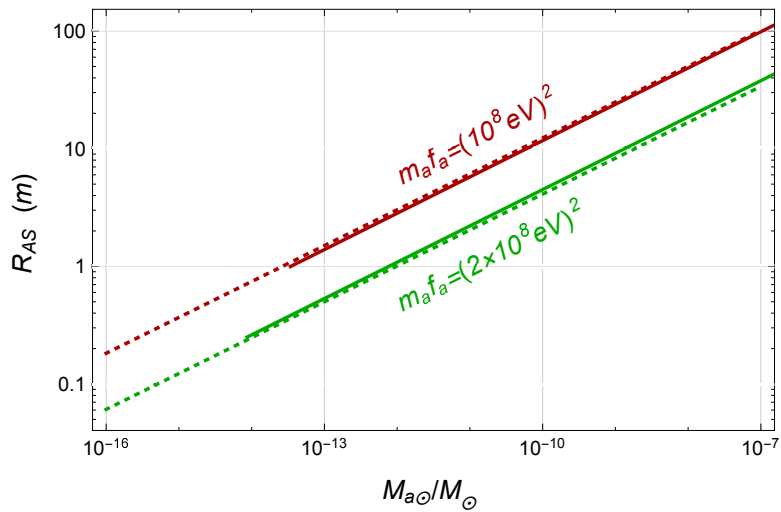}
\caption{Dense axion star radius as a function of its mass, as estimated using the Thomas-Fermi approximation. The solid lines are from numerical calculations and the dotted lines are from the fitted functions in Eq.~\eqref{eq:dense-axion-size}. }
\label{fig:axionsize}
\end{center}
\end{figure}

For a dense axion star, the self-interaction of axions is balanced by the gravitational attractive interaction. Following the calculation in Ref.~\cite{Braaten:2015eeu} with the Thomas-Fermi approximation, we show the dense axion star size and mass in Fig.~\ref{fig:axionsize}. We can fit the relation between the axion star radius and mass approximately as
\beqa
R^{\rm dense}_{\rm AS} = 1.5 \,\mbox{m}\, \times \,
\left( \frac{10^{11}~\mbox{GeV}}{f_a} \right)^{1/2}\,\left( \frac{10^{-4}~\mbox{eV}}{m_a} \right)^{1/2} \left( \frac{M_{a\odot}}{10^{-13}\,M_\odot} \right)^{0.3} \,,
\label{eq:dense-axion-size}
\eeqa
For the dense axion star, the energy density in the unit of $m_a^2 f_a^2$ can be order of unit. 
%
%

%

In our analysis later, we will treat $M_{a\odot}$, $m_a$, and $f_a$ as three independent model parameters. We note that depending on the detailed non-perturbative effects to generate the axion particle mass, there could be some model-dependent relation among these three parameters.

\section{Axion Electromagnetism}
\label{sec:axion-EM}
To observe the existence of the axion stars beyond its gravitational interaction~\cite{Kolb:1995bu,Fairbairn:2017dmf}, one could adopt other interactions of axion with SM particles. The relevant interactions are
\beqa
{\cal L} \supset \, -\frac{1}{4}F_{\mu\nu}F^{\mu\nu}\, - \,c_\gamma\, \frac{\alpha}{4\pi\,f_a}\,a\,F_{\mu\nu} \widetilde{F}^{\mu\nu}  - c_\psi \, \frac{\partial_\mu a}{f_a} \,\overline{\psi} \gamma^\mu \gamma^5 \psi \,.
\eeqa
Here, we use $\psi$ to represent SM fermions including electrons and quarks and $\widetilde{F}^{\mu\nu} = \frac{1}{2}\epsilon^{\mu\nu \rho\sigma}F_{\rho \sigma}$. With the existence of axion field, the Maxwell's equations are modified as
\beqa
\partial_\mu F^{\mu\nu} = j^\nu_\psi + j^\nu_a \equiv  \sum_\psi \, q_\psi \,e \,\overline{\psi} \gamma^\nu\,\psi - \frac{c_\gamma \,\alpha}{\pi\,f_a} \, \partial_\mu a \, \widetilde{F}^{\mu\nu} \,.
\eeqa
In terms of $\vv{B}$ and $\vv{E}$ fields, we have the axion-generated current as
\beqa
j^0_a = - \frac{c_\gamma \,\alpha}{\pi\,f_a}\,\nabla a \,\cdot\, \vv{B} \,, \qquad \mbox{and}\qquad 
\vec{j}_a = \frac{c_\gamma \,\alpha}{\pi\,f_a}\,\left[ (\partial_t a) \vv{B}\, +\, (\nabla a) \times \vv{E} \right] \,. 
\eeqa
Here, the axion field $a(\vec{x}, t)$ has variations in both time and space. In the Lorenz gauge with $\partial_\mu A^\mu=0$, the equation of motion of the gauge field is $\partial^2 A_\mu = j_\mu$. Using the Green's function method and taking the retarded boundary condition, we have the gauge field as~\cite{Jackson:1998nia}
\beqa
A_a^{\mu}(\vec{x},t) = \frac{1}{4\pi}\int d^3y \, \frac{ j_a^\mu(y, t - |\vec{x} - \vec{y} |) }{|\vec{x} - \vec{y}|} \,.
\label{eq:gauge-field-retarded}
\eeqa
If we ignore the time-dependence of the background $\vv{B}$ field and only keep the sinusoidal time dependence $e^{i\,m_a\,t}$ of the axion field, the gauge field is the real part of
\beqa
A_a^{\mu}(\vec{x},t) = \frac{e^{i\,m_a\,t}}{4\pi}\int d^3y \, \frac{j_a^\mu(y)\,e^{-i\,m_a\, |\vec{x} - \vec{y}|}}{|\vec{x} - \vec{y}|} 
\simeq e^{i\,m_a\,t}\, \frac{e^{-i\,m_a\,r}}{4\pi\,r}\int d^3y \, j_a^\mu(y)\,e^{i\,m_a\, \vec{n}_x\cdot \vec{y}} \,.
\label{eq:Aamu}
\eeqa
Here, we have made the approximation of $r\equiv |\vec{x}| \gg |\vec{y}|$ with $|\vec{x} - \vec{y}| \simeq r - \vec{n}_x \cdot \vec{y}$ and $\vec{n}_x \equiv \vec{x}/r = \hat{r}$. It is understood that the spacially-dependent current $j_a^\mu(y)$ has $\vec{j}_a(y)=\left[ m_a\,a(y) \vv{B} + \nabla a(y) \times \vv{E}\right]\,c_\gamma \alpha/(\pi f_a)$. 

In the case with a homogeneous and static magnetic field and a zero electric field, we have 
\beqa
A^0_a &=& -\frac{c_\gamma \alpha}{\pi f_a}\,  \frac{e^{i\,m_a(t -r)}}{4\pi\,r} \int d^3y\,\nabla a(y)\cdot \vv{B} \,e^{i\,m_a\, \vec{n}_x\cdot \vec{y}} \,,
\label{eq:A0-general} \\
\vv{A}_a &=& \frac{c_\gamma \alpha}{\pi f_a}\, \vv{B}\, \frac{m_a\,e^{i\,m_a(t -r)}}{4\pi\,r} \int d^3y\,a(y) \,e^{i\,m_a\, \vec{n}_x\cdot \vec{y}} \,.
\label{eq:Ai-general}
\eeqa
If the axion field is also homogeneous (at least in the region of large magnetic field) with $a(y)=a_0$, we have $A^0_a =0$. Using the first equality of Eq.~(\ref{eq:Aamu}) and changing the integrating variable, it is easy to show that $\vv{A}_a$ is independent of $\vec{r}$. So, there is no axion-generated magnetic field, and hence no radiated power. 

In reality, the axion field does not have correlation beyond its de Broglie wave length $d_a=1/(m_a \, v)$. As a result, we anticipate some inhomogeneity of the axion field and a nonzero $A^0_a$. For this general situation, there may not be correlation between the two directions of the generated $\vv{E}_a$ and $\vv{B}_a$. For a volume, $V$, of constant magnetic field and after (randomly) averaging out $V/d_a^3=V\,m^3_a v^3$ small correlated axion filed boxes (within its de Broglie wave length), we have the radiated power as
\beqa
&&\hspace{-0.5cm}\frac{dP}{d\Omega} = r^2\,\left|\vec{n}_x \cdot \vv{E}_a \times \vv{B}_a\right|
\approx \left(\frac{c_\gamma \alpha}{\pi f_a}\right)^2 \, \frac{m_a^4\,|\vv{B}|^2}{16\pi^2}\, \left(\frac{a}{m_a^3} \right)^2 \frac{V}{d_a^3} 
\,=\,\left(\frac{c_\gamma \alpha}{\pi f_a}\right)^2 \, \frac{|\vv{B}|^2}{16\pi^2}\, \left(\frac{\rho_a\,V\,v^3}{m_a} \right) 
\,, \\
&&\hspace{-0.7cm}\approx1.2\times 10^{-12}\,\mbox{W}\,\times\, c_\gamma^2 \left( \frac{10^{11}\,\mbox{GeV}}{f_a}\right)^2 \,\left( \frac{10^{-4}\,\mbox{eV}}{m_a}\right) \, \left( \frac{|\vv{B}|}{10^{10}\,\mbox{G}}\right)^2\, \left[ \frac{V}{(10\,\mbox{km})^3}\right]\,\left(\frac{v}{10^{-3}}\right)^3\, \left( \frac{\rho_a}{0.3\,\mbox{GeV}/\mbox{cm}^3}\right) \,, \nonumber
\eeqa
which is very small for a very far-away source like a neutron star. 

For the axion dark matter in our galaxy, the large volume effect could enhance the flux, although the corresponding averaged magnetic field is much smaller, $|\vv{B}|_{\rm galaxy} \sim \mathcal{O}(10\,\mu G)$. Following Ref.~\cite{Haverkorn:2014jka}, we treat the magnetic field as a constant one within a simple cylinder-shape with a radius of 20~kpc and half-height of 2~kpc. For the dark matter distribution in our Milky Way Galaxy, we use the isotropic Einasto profile~\cite{Graham:2006ae} with 
\beqa
\rho^{\rm MW}_a(r) = \rho_{\odot} \, \mbox{exp} \left( - \frac{2}{\beta} \left[ \left( \frac{r}{r_s} \right)^\beta - \left(\frac{r_\odot}{r_s} \right)^\beta \right] \right) \,,
\eeqa
with $\rho_\odot = 0.4~\mbox{GeV}/\mbox{cm}^3$, $r_s=20$~kpc, $r_\odot = 8.5$~kpc and $\beta= 0.17$. The spectral flux density that can be observed by a detector on the Earth is 
\beqa
S &=& \left(\frac{c_\gamma \alpha}{\pi f_a}\right)^2 \, \frac{v^3}{16\pi^2\,m_a^2} \int dV\, \frac{\rho^{\rm MW}_a(r) |\vv{B}(\vec{r})|^2}{d(\vec{r})^2} \sim  \left(\frac{c_\gamma \alpha}{\pi f_a}\right)^2 \, \frac{v^3}{16\pi^2\,m_a} \rho_{\odot} |\vv{B}_{\rm galaxy}|^2\, \times\,2.6~\mbox{kpc}  \\
&=& 5.7 \times 10^{-45}\,\frac{\mbox{W}}{\mbox{m}^2\,\mbox{Hz}} \,\times \, c_\gamma^2 \left( \frac{10^{11}\,\mbox{GeV}}{f_a}\right)^2 \,\left( \frac{10^{-4}\,\mbox{eV}}{m_a}\right)^2 \, \left( \frac{|\vv{B}|_{\rm galaxy}}{10\,\mu \mbox{G}}\right)^2\, \left(\frac{v}{10^{-3}}\right)^3\, \left( \frac{\rho_{a\odot}}{0.4\,\mbox{GeV}/\mbox{cm}^3}\right) \,. \nonumber 
\eeqa
Here, the distance of a source to the observer is $d(\vec{r})^2 = r^2 + r_\odot^2 - 2\,r\,r_\odot \,\cos{\theta}$, with $\theta$ as the angle between $\vec{r}$ and the direction from the galactic center to the Sun. In terms of the unit of Jansky with $1\,\mbox{Jy} = 10^{-26}\mbox{W}/(\mbox{m}^2\,\mbox{Hz})$, the total spectral flux density is at the order of $10^{-19}\,\mbox{Jy}$ and very small (see Refs.~\cite{Pshirkov:2007st,Kelley:2017vaa,Sigl:2017sew} for similar calculations).

\section{Power of Axion Star in Static Magnetic Field}
\label{sec:radiation}
\subsection{Without Medium}
\label{sec:no-medium}
For the axion star, there are two effects that can potentially enhance the radiation power. One is the much higher energy density carried by the axion star. The second is that being a Bose-Einstein Condensation (BEC) state, the axion star is more compact and potentially has a larger contribution to radiation power from different regions. To simplify our discussion, we will simply assume the spacial wave-function of the axion star follows the $1S$-state of the hydrogen atom.~\footnote{The spacial profile of a dense axion is different from the $1S$-state one, but does not change much our later order-of-magnitude estimation.} So, the axion star field value is
\beqa
a(\vec{r}) =\frac{\sqrt{M_{a\odot}}}{m_a}\, \frac{1}{\sqrt{4\pi}}\, \frac{2}{R_{\rm AS}^{3/2}}e^{-r/R_{\rm AS}} \,,\label{eq:axion-star-solution}
\eeqa
where we have chosen the normalization to have the total axion-star mass to be $M_{a\odot}\approx {\cal N}_a \,m_a$ in the non-relativistic approximation. Performing a straight-forward integrations of Eqs.~(\ref{eq:A0-general})(\ref{eq:Ai-general}) and in the limit of $R_{\rm AS} \gg 1/m_a$, we have the field value as
\beqa
A^0_a &=&-\frac{c_\gamma \alpha}{\pi f_a} \,\frac{e^{i m_a (t - r)}}{\sqrt{4\pi}\,r} \, \frac{\sqrt{M_{a\odot}}}{m_a^4\,R_{\rm AS}^{5/2}}\, 4\,i\,\hat{r}\cdot \vv{B} \equiv -f_1(r)\,i\,B\cos{\theta} \,, \\
\vv{A}_a &=& \frac{c_\gamma \alpha}{\pi f_a} \,\frac{e^{i m_a (t - r)}}{\sqrt{4\pi}\,r} \, \frac{\sqrt{M_{a\odot}}}{m_a^4\,R_{\rm AS}^{5/2}}\,4\,\vv{B}  \equiv f_1(r) \vv{B}\,.
\eeqa
Choosing $\vv{B}$ along the $\hat{z}$ direction, we have $\hat{r}\cdot \vv{B} = B \cos{\theta}$. Using the relation of $\nabla \times [ f_1(r) \vv{B}] = [\nabla f_1(r)] \times \vv{B} = f'_1(r) \hat{r} \times \vv{B}$ for a constant $\vv{B}$, we have the generated $\vv{B}_a \approx  -i m_a f_1(r) \, \hat{r} \times \vv{B}$ at the leading order of $1/r$. The generated electric field $\vv{E}_a \approx - m_a\,B\,\cos{\theta} \,f_1(r)\,\hat{r} + \frac{1}{r}\,i \,B\,\sin{\theta}\,f_1(r) \hat{\theta} - i m_a\,f_1(r) \vv{B}$. The radiated power is 
\beqa
\frac{dP}{d\Omega} &=& r^2\,\left|\hat{r} \cdot \vv{E}_a \times \vv{B}_a\right|  \approx r^2 \,m_a^2 \, f_1^2(r)\, B^2 \sin^2{\theta}  = \left(\frac{c_\gamma \alpha}{\pi f_a}\right)^2 \, \frac{4}{\pi}\, \frac{M_{a\odot}}{m_a^6\,R^5_{\rm AS}}\,|\vv{B}|^2\,\sin^2{\theta} \,, \label{eq:power-no-medium}\\
&=& 4.2 \times 10^{11}\,\mbox{W}\times c_\gamma^2\,\sin^2{\theta}\, \left( \frac{10^{11}\,\mbox{GeV}}{f_a}\right)^2 \,\left( \frac{10^{-4}\,\mbox{eV}}{m_a}\right)^6 \, \left( \frac{|\vv{B}|}{10^{10}\,\mbox{Gauss}}\right)^2\, \left( \frac{M_{a\odot}}{10^{-13}M_\odot}\right)\, \left( \frac{1.5\,\mbox{m}}{R_{\rm AS}}\right)^5 \, , \nonumber\\  
&=& 4.2 \times 10^{11}\,\mbox{W}\times c_\gamma^2\,\sin^2{\theta}\, \left( \frac{m_a\,f_a}{(10^8\,\mbox{eV})^2}\right)^{1/2} \,\left( \frac{10^{-4}\,\mbox{eV}}{m_a}\right)^4 \, \left( \frac{|\vv{B}|}{10^{10}\,\mbox{Gauss}}\right)^2\, \left( \frac{M_{a\odot}}{10^{-13}M_\odot}\right)^{-0.5} \,,
 \nonumber 
\eeqa
in the last step, we have used the relation in Eq.~\eqref{eq:dense-axion-size} for the radius of a dense axion star.
The magnetic field could change its direction, so there could be some source-dependent frequency for $\theta$.  Numerically, the time scale of the collision is $\mathcal{O}(\mbox{few}\,R_{\rm NS}/v_{\rm rel}) \approx \mathcal{O}(0.1\,\mbox{s})$ while the period of rotation of the typical pulser is $10^{-3}$--$10^1$ s. This implies that the radio wave power from the collision may oscillate in time.
This observation may help to disentangle signals from backgrounds, if the power is large enough. For the source at the scale of 1\,pc, with $B=10^{10}$\,Gauss and for $m_a = 10^{-4}$\,eV, the estimated spectral flux density is around $10^{-7}$\,Jy. 

\subsection{With Medium}
\label{sec:medium}
If there is a medium surrounding the axion star and with a static magnetic field, the response of medium to the axion-generated electric field may generate a large radiation power. For a plasma of electrons and ions, the electron motions responding to the oscillating electric field behave as an oscillating electric dipole moment~\cite{Hill:2016zos}. The large number density of electrons could potentially increase the radiation power. However, for a macroscopic medium, the induced electric field from the axion and static magnetic field is also suppressed by the plasma frequency, so the axion-star radiated power is not amplified in the medium. 

For macroscopic material, one need to calculate the applied electric field or the polarization field $\vv{P}$, which is related to the electric field as $\vv{P}=\chi\vv{E}$ with $\chi$ as the electric susceptibility or the electric flux densities $\vv{D}= \epsilon\,\vv{E}$ with $\epsilon = 1 + \chi$ (we have chosen vacuum permittivity $\epsilon_0=1$ in the natural unit). Based on the simple Drude model, one can calculate the frequency-dependent permittivity as
\beqa
\epsilon(\omega) =  1 + \chi(\omega) =  1 + \frac{\omega_p^2}{\omega_0^2 - \omega^2 + i \,\gamma\,\omega} \,,
\eeqa
where $\omega$ is the frequency of the applied electric field; $\gamma$ is a measure of the rate of collisions per unit time and for a typical conductor $\gamma = \mathcal{O}(10^{14}\,\mbox{s}^{-1})=\mathcal{O}(0.066\,\mbox{eV})$;  the oscillator frequency $\omega_0=0$ for free electrons and non-zero for bounded electrons; the plasma frequency $\omega_p=(n_e\,e^2/m_e)^{1/2}$.  

Inside the medium and neglecting spatial derivative terms, the relevant Maxwell equations have~\footnote{We have assumed  that the axion star equation of motion receives small corrections from the external magnetic and electric field, which is a good approximation.}
\beqa
\nabla \cdot \vv{E}_a &=& j^0 = - \nabla \cdot \vv{P}_a  - \frac{c_\gamma\,\alpha}{\pi \, f_a}\,\nabla a \cdot \vv{B} \, , \\
\nabla \times \vv{B}_a - \frac{\partial \vv{E}_a}{\partial t}  &=& \vv{j}  = \frac{\partial \vv{P}_a }{\partial t} + \frac{c_\gamma\,\alpha}{\pi \, f_a}\, (\partial_t a) \vv{B} \,.
\eeqa
Solving the Maxwell equation, the effective electric field generated by the axion field around the source region is
\beqa
\vv{E}^{\rm source}_a \approx - \frac{1}{1 + \chi}\, \frac{c_\gamma \,\alpha}{\pi\,f_a}\,a\,\vv{B} \,.
\eeqa
The corresponding axion-generated current in the medium has 
\beqa
j^0_a = - \left( 1 - \frac{\chi}{1 + \chi} \right)  \frac{c_\gamma\,\alpha}{\pi \, f_a}\,\nabla a \cdot \vv{B} \,, \qquad \qquad 
\vec{j}_a = \left( 1 - \frac{\chi}{1 + \chi} \right)  \frac{c_\gamma\,\alpha}{\pi \, f_a} \,  (\partial_t a) \vv{B}  \,.
\eeqa
Similar to Eqs.~(\ref{eq:A0-general})(\ref{eq:Ai-general}), we can use the above medium-responded current to calculate the radiation power. It is clear that if the medium has a large value of $\chi \gg 1$, the radiated power is suppressed by $1/\chi^2$. Therefore, we conclude that the medium can not amplify the radiation power.~\footnote{The medium effects ware not properly taken into account in Ref.~\cite{Iwazaki:2014wta,Iwazaki:2015zpb}.}

Around the surface region of the neutron star, the plasma frequency is $\omega_p \approx 1.1 \times 10^5$~eV for the nuclei (iron) density of $8\times 10^6$\,g/cm$^3$~\cite{Chamel:2008ca}. For the oscillation frequency and using the radial potential of $V = - Q_{\rm ion}\,\alpha/r + l^2/(2 \,m_e \,r^2)$ with $l$ as the angular momentum number, we can have 
\beqa
\omega_0 \approx \frac{\alpha^2\,Q^2_{\rm ion}\,m_e}{l^3} \approx 6.8 \times 10^5~\mbox{eV} \,.
\eeqa
For $l=1$, we have $|\chi| \approx \omega_p^2/\omega_0^2 \approx  0.03$. The medium effect for the axion-star-radiated power could be small, even when the axion star is actually overlapping with the neutron star. The radiated power in Eq.~(\ref{eq:power-no-medium}) without medium can be trusted as an order-of-magnitude estimation. 

Before we end this section, we also comment on the possibility of using neutron electric dipole moment in Ref.~\cite{Raby:2016deh}. We show that there is no volume-square or $N_n^2$ ($N_n$ is the number of neutrons) effect from the integration or interference of different regions, as can be seen from Eq.~\eqref{eq:gauge-field-retarded}. As a result, the neutron-medium-generated power is negligible. For completeness, we provide the order-of-magnitude estimation of the power from oscillating electric dipole moment of the neutron. In the background of classical axion field, the neutron electric dipole is~\cite{Pospelov:1999ha,Graham:2013gfa}
\beq
\vec{d}(\vec{r},t)\simeq {e\over m_N}{a(\vec{r},t)\over f_a}=: \vec{d}_0 {a(\vec{r})\over f_a}\sin(m_a t) \,,
\qquad 
|\vec{d}_0|\simeq2.4\times10^{-16}e\cdot{\rm cm} \,,
\eeq
where $a(\vec{r})$ is given by Eq.~\eqref{eq:axion-star-solution}. Following the same calculation as before, we have the radiated power estimated as
\beqa
{dP\over d\Omega} &\sim& \frac{\alpha^2\,m_a^4\,|\vec{d}_0|^2}{f_a^2} \, \left| \int \,d^3y\,n_n(\vec{y}) e^{i\,m_a\,\vec{n}_x\cdot\vec{y}}\,a(\vec{y})\right|^2  \approx  \frac{\alpha^2\,|\vec{d}_0|^2\,\bar{n}_n^2\,M_{a\odot}}{f_a^2\,m_a^6\,R_{\rm AS}^5} \, \\
&=& 5.4\times10^{-13}\,\mbox{W}\left( \frac{10^{11}\,\mbox{GeV}}{f_a}\right)^2 \,\left( \frac{10^{-4}\,\mbox{eV}}{m_a}\right)^6 \, \left( \frac{M_{a\odot}}{10^{-13}M_\odot}\right) \, \left( \frac{\bar{n}_n}{10^{27}\,\rm{cm}^{-3}}\right)^2\, \left( \frac{1.5\,\mbox{m}}{R_{\rm AS}}\right)^5 \,.
\label{eq:dense-axion-radiation-power}
\eeqa
Here, $n_n$ is the number density of the neutron and $\bar{n}_n$ is the averaged number density. In our estimation, we have assumed $m_a^{-1}\ll R_{\rm AS}\lesssim R_{\rm NS}$. Because of the interference effects, the total power is suppressed by a factor of $1/(m_a R_{\rm AS})^8$ on top of $N_n^2 \approx (\bar{n}_n R^3_{\rm AS})^2$. Compared to the power generated from interacting with the magnetic field in Eq.~\eqref{eq:power-no-medium}, the power generated from oscillating electric dipole moment of neutrons is negligible.  

\section{Signal when Axion Stars Encounter Neutron Stars}
\label{sec:neutron-star}
After previous calculations, we have learned that the largest radiation power for the axion star can be obtained if it passes through a large magnetic field. Neutron stars are the obvious objects, but one could also consider other objects like white dwarfs with a smaller magnetic field but a larger population. Therefore, in this section we will estimate the encounter rates as well as the experimental detectability. 

\subsection{Encounter rate}
\label{sec:encoutner}
Assuming that the total energy of axion stars is comparable to the total dark matter energy, the total number of axion stars in our galaxy is roughly $2\times 10^{24}\times[10^{-13}\,M_{\odot}/M_{a\odot}]$. Around our solar system, the axion-star density is around $n_{\rm AS}\approx f_{\rm AS}\times1.1\times 10^{11}\,\mbox{pc}^{-3} \,[10^{-13}\,M_{\odot}/M_{a\odot}]$ with $f_{\rm AS}$ as the fraction of axion-star energy in total dark matter energy.~\footnote{Depending on the axion-star mass, the mirco-lensing-based searches for MACHO by Subaru Hyper Suprime-Cam in Ref.~\cite{Niikura:2017zjd} has imposed a constraint on $f_{\rm AS}$ to be below a few percent for $M_{a\odot}=10^{-12}M_\odot$ to no constraint for $M_{a\odot}=10^{-14}M_\odot$.}  In the meanwhile, there are around $10^9$ neutron stars in our galaxy with the magnetic field ranging from $10^8$ to $10^{15}$ Gauss~\cite{Shapiro:1983du}. Even slightly outside the neutron star, there is still a large magnetic field. Therefore, we do not require that the axion star strictly collides with the neutron star, but calculate the rate of the event where the minimal distance between the axion star and the neutron star is $(R_{\rm AS}+\ell\,R_{\rm NS})$ with $\ell \ge 1$. To estimate the encounter rate, we simply use the geometric cross section augmented by the ``gravitation focus" effect or the increasing Safronov number~\cite{1972epcf.book.S}. The cross section has the form of 
\beqa
\sigma\,=\,\pi\,(R_{\rm AS}+\ell\,R_{\rm NS})^2\,\left[ 1 + \frac{v^2_{\rm esc}}{v^2_{\rm rel}} \right] =\pi\,(R_{\rm AS}+\ell\,R_{\rm NS})^2\left[1+{2\,G_N\,(M_{\rm NS}\,+ M_{a\odot})\,\over v^2_{\rm rel} \,(R_{\rm AS}+\ell\,R_{\rm NS} )}\right] \,,
\eeqa
with $R_{\rm NS} \sim 10$~km and $M_{\rm NS}\sim M_{\odot}$ as the neutron-star radius and mass, respectively. The relative speed is approximately the averaged axion-star speed in our galaxy and around $300\,\mbox{km}/\mbox{s}$. The escaping velocity is estimated to be $v_{\rm esc} \approx 1.6\times 10^5\,\ell^{-1/2}\,\mbox{km}/\mbox{s}$. The encounter events per year are estimated to be
\beqa
N_{\rm collision}/ {\rm year} &=& n_{\rm AS}\times n_{\rm NS}\times \sigma v\times V_{\rm galaxy}
\simeq n_{\rm AS}\times N_{\rm NS} \times \sigma v_{\rm rel}  \\
&=&0.003 \times \ell \, f_{\rm AS} \, \left(\dfrac{10^{-13}\,M_\odot}{M_{a\odot}}\right)\, \left(\frac{N_{\rm NS}}{10^9} \right)\,.
\label{eq:collision-rate}
\eeqa
For a lighter axion star, the encountering events can happen more frequently in our galaxy. From Eq.~\eqref{eq:collision-rate}, the encounter events per year are not large. However, we note that the calculation here is just the order-of-magnitude estimation. The actual encounter rate could be increased significantly if a dark matter clump happens in the neutron-star-rich region. As can be seen from Eq.~\eqref{eq:collision-rate}, the encounter rate can become large with a large value of $\ell$, at the price of a suppressed magnetic field by $\ell^{-3}$.~\footnote{Here, it is assumed that the neutron star is the point-like magnetic dipole source outside the neutron star. For the realistic magnetic field, the corresponding power could be different.} Furthermore, if the number of neutron stars is larger than $10^9$ in our galaxy, then we may also obtain a larger encounter rate. Having these possibilities in mind, we will simply assume that encounter event can happen at the scale of one year and discuss the detectability. 

From Eq.~\eqref{eq:power-no-medium} and when one axion star enters the vicinity of a neutron star, the radiated power can vary from $10^{11}$~W to $10^{21}$~W, depending on the magnetic field size from $10^{10}$~Gauss to $10^{15}$~Gauss. This kind of event is also likely to be a transient one with the time scale of $\mathcal{O}(\mbox{few}\,R_{\rm NS}/v_{\rm rel}) \approx \mathcal{O}(0.1\,\mbox{s})$. Since this encounter event is like a point source from the  astrophysical point of view, the closest event to our solar system may have the highest probability to be detected. For the $\sim 10^9$ neutron stars in the our galaxy disk, the averaged density is roughly one per parsecs. For sure, the anticipated encounter rate at a distance of parsecs is very tiny. 

Using Eq.~\eqref{eq:power-no-medium} and assuming a typical source distance of 1~kpc, the estimated spectral flux density from axion--neutron-star encounter is
\beqa
S \approx 2.9\times10^{-13}\,\mbox{Jy}\times \left( \frac{m_a\,f_a}{(10^8\,\mbox{eV})^2}\right)^{1/2} \,\left( \frac{10^{-4}\,\mbox{eV}}{m_a}\right)^5\, \left( \frac{10^{-13}M_\odot}{M_{a\odot}}\right)^{0.5} \, \left( \frac{|\vv{B}|}{10^{10}\,\mbox{Gauss}}\right)^2\, \left( \frac{1\,\mbox{kpc}}{d_{\rm source}}\right)^2  \,,
\label{eq:spectral-flux-density}
\eeqa
which is applicable for the dense axion-star mass above the upper bound for the diluted axion mass in Eq.~\eqref{eq:dilute-condition}. The encounter rate is given in Eq.~\eqref{eq:collision-rate} with the typical source distance at kpc scale. The spectral density flux highly depends on the axion particle mass, the neutron-star magnetic field and the location of the encountering event.

\subsection{Detectability }
\label{sec:detectability}
The signal from axion conversion to photon has the radio frequency peaked at $\nu = m_a/2\pi = 24\,\mbox{GHz}\times (m_a/10^{-4}\,\mbox{eV})$. Both the axion-star spacial profile and the inhomogeneity of the magnetic field can generate a non-zero but small bandwidth. For the interesting mass range, $10^{-6}$--$10^{-3}$\,eV, of QCD axion, the corresponding frequency is 240~MHz to 240~GHz. It is intriguing to see if the on-going or future radio telescope can have chance to discover axion-neutron-star encounter event.  

The existing Arecibo Observatory has a 305-meter single-aperture telescope with covered frequencies from 300~MHz to 10~GHz~\cite{Giovanelli:2005ee}. For instance, some fast radio burst was discovered by this radio telescope~\cite{Spitler:2014fla}. The sensitivity to the continuous spectral flux density is around $0.89\,\mu{\rm Jy}\,({\rm hr}/t_{\rm obs})^{1/2}$ with $t_{\rm obs}$ as the observing time and the band width of $\mathcal{B}=1000$~MHz. The existing Green Bank Telescope (GBT) has a 100 by 110 meter active surface and can cover the frequency range of 290~MHz--115.3~GHz~\cite{GBT}. The sensitivity is $5.89\,\mu{\rm Jy}\,({\rm hr}/t_{\rm obs})^{1/2}$ for the bandwidth of 400~MHz. The existing Karl G. Jansky Very Large Array (JVLA) has 27 independent antenna, each of which has a dish diameter of 25 meters, and can cover the frequency range of 1--50~GHz~\cite{JVLA}. The sensitivity is $5.89\,\mu{\rm Jy}\,({\rm hr}/t_{\rm obs})^{1/2}$ for the bandwidth of 400~MHz. The Five hundred meter Aperture Spherical Telescope (FAST)~\cite{Nan:2011um}, located in Guizhou, China, can cover the frequency range of 70~MHz--3~GHz (up to 8~GHz from future upgrading). The sensitivity on the spectral flux density is around $0.92\,\mu{\rm Jy}\,({\rm hr}/t_{\rm obs})^{1/2}$ with the bandwidth of 800~MHz. The Square Kilometre Array (SKA) (to be constructed) has the frequency from $50$\,MHz to $350$\,MHz (SKA-low) and $350$\,MHz to $14$\,GHz (SKA-mid). The sensitivity can be reached to be $3.36\,\mu{\rm Jy}\,({\rm hr}/t_{\rm obs})^{1/2}$ with $\mathcal{B}=300$~MHz (SKA-low)  and $0.75\,\mu{\rm Jy}\,({\rm hr}/t_{\rm obs})^{1/2}$ with $\mathcal{B}=770$~MHz (SKA-mid)~\cite{SKA}.

To estimate the sensitivity of different telescopes on the axion-star generated power, we can convert the spectral flux density in Eq.~\eqref{eq:spectral-flux-density} to the brightness temperature~\cite{RadioAstronomy}
\beqa
T_{\rm AS} \equiv \frac{1}{2} \, A_{\rm eff}\, S \, = \, 0.36\, \mbox{K}\,\times\,\left( \frac{A_{\rm eff}}{10^3 \, \mbox{m}^2} \right) \left( \frac{S}{\mbox{Jy}}\right)\,.
\label{eq:temperature-S}
\eeqa
Here, the factor of $1/2$ takes into account that the telescope usually only receives one polarization; the parameter $A_{\rm eff}$ is the effective area of the telescope. Depending on the telescope system temperature $T_{\rm sys}$ (typically order of 10~K), which includes both sky noise and instrumental noise,~\footnote{
The continuous spectrum of the background neutron star noise is also included in $T_{\rm sys}$. The expected flux density is $10^{-3}$--$100$\,Jy~\cite{Lyutikov:2002kh,Camilo:2006zf}, which is at most order of $10$\,K in terms of the brightness temperature.} the minimal detectable brightness temperature is roughly 
\beqa
T_{\rm min} =\sqrt{2}\, \frac{T_{\rm sys}}{\sqrt{\mathcal{B}\,t_{\rm obs}}} \,,
\label{eq:temperature-min}
\eeqa
Combining Eqs.~\eqref{eq:temperature-S} and \eqref{eq:temperature-min}, the rough sensitivity on the spectral flux density is 
\beqa
S_{\rm min}  
= 2.1 \, \mu\mbox{Jy}\,  \times\,\left(\frac{10^3\,\mbox{m}^2/\mbox{K} }{A_{\rm eff}/T_{\rm sys}}\right) \left(\frac{1000~\mbox{MHz}}{\mathcal{B}}\right)^{1/2} \,\left(\frac{1\,\mbox{hr}}{t_{\rm obs}}\right)^{1/2} \,.
\eeqa
For different telescopes, one can find a summary of different $A_{\rm eff}/T_{\rm sys}$ in Ref.~\cite{SKA}. 

\begin{figure}[th!]
\begin{center}
\includegraphics[width=0.65\textwidth]{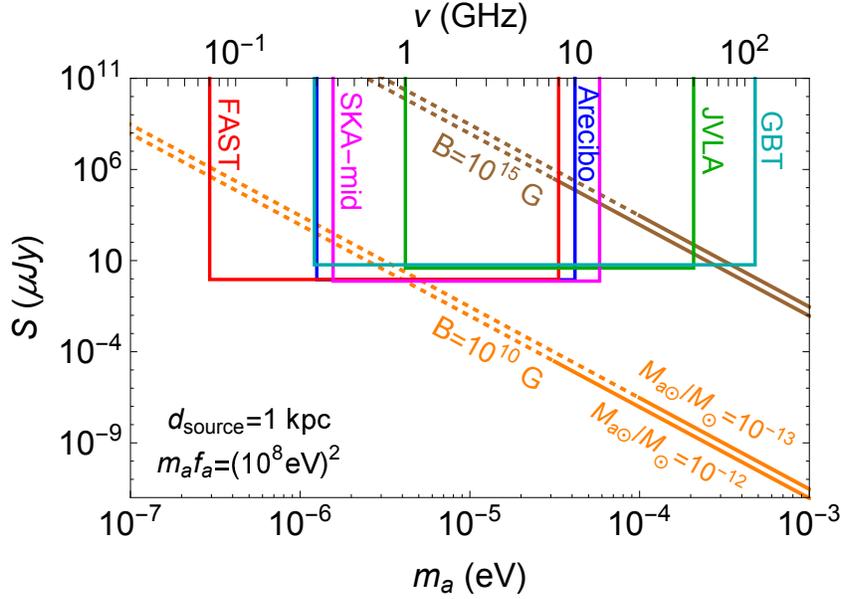}
\caption{The detectability of the radio signal from the encounter event of the neutron star and density axion star. The sensitivities from Arecibo, GBT, JVLA, FAST, and SKA-mid are shown for $t_{\rm obs}=1$\,hour. The lower end of the orange and brown lines correspond to the lower bound on the dense axion star mass, see Eq.~\eqref{eq:dilute-condition}. }
\label{fig:detection}
\end{center}
\end{figure}

In Fig.~\ref{fig:detection}, we show the dense-axion-star generated radiation power when it encounters neutron stars with a magnetic field of $10^{10}$\,Gauss and $10^{15}$\,Gauss, for different axion particle masses. We fix the product of $m_a f_a = (10^8\,\mbox{eV})^2$ to approximately match the QCD axion relation. For a large value of magnetic field above $\approx 10^{13}$~Gauss, the sensitivities of radio telescopes are good enough to potentially observe the transient $\mathcal{O}(0.1\,\mbox{s})$ radio source, very similar to the fast burst radio, except located within our galaxy.

\section{Discussion and Conclusions}
\label{sec:conclusion}
So far, we have concentrated on the radiation power from a dense axion star. Our calculation has been kept very general such that our Eq.~\eqref{eq:power-no-medium} can be applicable to other axion stars like a diluted axion star. For sure, the diluted-axion-star radius $R_{\rm AS}$, given by the Eq.~\eqref{eq:dilute-axion-size}, is much larger than the one for a dense axion star in  Eq.~\eqref{eq:dense-axion-size}. Since the flux is proportional to $R_{\rm AS}^{-5}$ as shown in Eq.~\eqref{eq:dense-axion-radiation-power}, the signal of a diluted axion star is dramatically suppressed compared with that of a dense axion star and is unlikely to be observed. This large suppression factor is due to the interference effect when one integrates the radiation power generated from different regions of the axion star. 

We also comment on the case of the non-QCD axion. In this case, the relation of $m_a f_a\approx (10^8\,{\rm eV})^2$ does not need to be satisfied. Requiring the symmetry breaking scale below the end-of-inflation scale and satisfying the astrophysical bound, the decay constant should be still in the range of $10^8\,\mbox{GeV} < f_a < 10^{13}\,\mbox{GeV}$. Keeping the axion star mass around $10^{-13}M_\odot$ and from Eq.~\eqref{eq:dilute-condition}, one could have $f_a = 10^{8}$\,\mbox{GeV} to saturate the astrophysical bound. To have a dense axion star, the axion particle mass is bounded from below $m_a \gtrsim 10^{-7}$\,eV. The maximum power for a dense non-QCD axion star could be 9 orders of magnitude higher than a dense QCD axion star, as can be seen from Eq.~\eqref{eq:power-no-medium}. Therefore, the radio telescopes could have a good coverage for some non-QCD dense axion stars if they search for spectral-line signals with a wide range of frequency, order of $10^{-2}$--$10^2$\,GHz. 

The transient radio signal when an axion star meets a neutron star or other astrophysical object is very similar to the FRB. Both of them have the signal lasting below one second. So far, the discovered FRB's have their origin at an extra-galactic distance~\cite{Katz:2016dti}. The axion-neutron-star encounter event has a smaller radiation power than a FRB, and is likely in the nearby region of our galaxy if discovered. Therefore, a future discovery of a galactic FRB should deserve a study of its spectrum to see if it has a clear spectra line peaked at the preferred QCD or other non-QCD axion mass. 

As shown in Eq.~\eqref{eq:collision-rate}, the axion-neutron-star encounter rate is not that high because of the small geometric size and population of neutron stars. An axion star could just enter the vicinity of a neutron star and could still generate a large enough radiation power. Far away from the neutron star, the magnetic field decreases as $\propto r^{-3}$ if we treat the neutron star as a point-like magnetic dipole source. The encounter rate increases as $\propto r$ while the spectral flux density decreases as $\propto r^{-6}$. For sure, we have not considered the situation with some local clump of axion stars in our galaxy disk where most of neutrons sit, which will possibly increase our estimation of the encounter rate. 

Other than neutron stars, one could also consider other astrophysical objects with a large magnetic field. White dwarfs are the obviously next candidate. In our galaxy, we anticipate $\mathcal{O}(10^{11})$ white dwarfs, which has their radius of $\mathcal{O}(10^4\,\mbox{km})$ and magnetic field of $10^6$--$10^9$\,Gauss~\cite{Ferrario:2015oda,2016MNRAS.455.3413K}. Using Eq.~\eqref{eq:collision-rate}, the encounter rate could be at the order of $100\,\ell\,f_{\rm AS}/\mbox{year}$. Because of its small magnetic field, the encounter event for a dense QCD axion star can be observed by radio telescopes only if it happens in a very nearby region, $\mathcal{O}(1\,\mbox{pc})$ from our solar system. Finally, the dense axion star may also enter our solar system and has an encounter event with our Sun. Both the generated radiation power and the encounter rate are too small to be observed.

In summary, we have estimated the radiation power when a dense axion star encounters a neutron star or other astrophysical objects with a large magnetic field. The spectral flux density is at the order of $\mu$Jy for a neutron star with $B\sim 10^{13}$~Gauss and the encounter event happening within one kpc from our solar system. The existing Arecibo, GBT, JVLA and FAST and the ongoing SKA radio telescopes have excellent discovery potential for the dense axion stars.

\subsection*{Acknowledgments}
We would like to thank Josh Berger for discussion and comments. This work is supported by the U. S. Department of Energy under the contract DE-FG-02-95ER40896. The work of YH is supported in part by the Grant-in-Aid for JSPS Fellows No.16J06151.

\providecommand{\href}[2]{#2}\begingroup\raggedright\endgroup

 \end{document}